\documentclass[sigconf]{acmart}
\setcopyright{none}
\usepackage{multirow}
\usepackage{algorithm,algpseudocode}
\usepackage{tikz}
\usetikzlibrary{positioning, arrows.meta, shapes.geometric, calc, backgrounds, fit, shadows}
\usepackage{booktabs}
\usepackage{setspace}
\usepackage{enumitem}
\usepackage{xurl}
\usepackage{graphicx}
\usepackage[normalem]{ulem}
\usepackage{xcolor}

\setlist[itemize]{leftmargin=*,topsep=0pt,itemsep=0pt,parsep=0pt}

\renewcommand\footnotetextcopyrightpermission[1]{}
\renewcommand{\shortauthors}{}%
\AtBeginDocument{
  \fancyhead{}
}
\acmYear{}
\acmDOI{}
\acmISBN{}
\AtBeginDocument{%
  \providecommand\BibTeX{{%
    \normalfont B\kern-0.5em{\scshape i\kern-0.25em b}\kern-0.8em\TeX}}}

\copyrightyear{}
\acmYear{}
\setcopyright{none}
\acmConference{}{}{}
\acmBooktitle{}
\acmPrice{}
\acmDOI{}
\acmISBN{}

\begin{document}

\title{Automated QoR improvement in OpenROAD with coding agents}

\author{Amur Ghose, Junyeong Jang, Andrew B. Kahng, Jakang Lee \\
\{aghose, juj015, abk, jal216\}@ucsd.edu \\
UC San Diego, La Jolla, California, USA}

\renewcommand{\shortauthors}{Ghose, et al.}

\begin{abstract}
EDA development and innovation has been constrained by scarcity of expert engineering resources. 
While leading LLMs have demonstrated excellent performance in coding and scientific reasoning tasks, their capacity to advance EDA technology itself has been largely untested. We present 
{\bf AuDoPEDA}, an autonomous, repository-grounded coding system built atop OpenAI models and a 
Codex-class agent that reads OpenROAD, proposes research directions, expands them into 
implementation steps, and submits executable diffs. Our contributions include (i) a closed-loop 
LLM framework for EDA code changes; (ii) a task suite and evaluation protocol on 
OpenROAD for PPA-oriented improvements; and (iii) end-to-end demonstrations 
with minimal human oversight. Experiments in OpenROAD achieve routed 
wirelength reductions of up to 5.9\%, effective clock period reductions of up to 10.0\%, and power reductions of up to 19.4\%.
\end{abstract}


\maketitle

\section{Introduction}
VLSI physical design (PD) remains heavily constrained by the availability of senior engineers who can reason across large, multi-language electronic design automation (EDA) codebases and long, iterative flows. Recent code-focused large language models (LLMs) show strong performance on local programming tasks, but their effectiveness degrades when the required context spans many files, undocumented invariants, and heterogeneous toolchains. The open-source OpenROAD project \cite{openroad,openroad-github} exemplifies this setting: it comprises millions of lines of C++, Tcl, Python, and Verilog, evolves on a weekly cadence, and offers only sparse cross-module documentation with many implicit interfaces. These characteristics make onboarding difficult for humans and pose an even greater challenge for autonomous coding agents.

\begin{figure}[t]
    \centering
    \begin{tikzpicture}[
        node distance=0.6cm,
        stage/.style={
            rectangle,
            rounded corners=2pt,
            draw=black!70,
            fill=black!3,
            minimum width=6.5cm,
            minimum height=0.9cm,
            align=left,
            inner xsep=6pt,
            font=\sffamily\footnotesize
        },
        arrow/.style={-Stealth, thick},
        note/.style={font=\sffamily\scriptsize, align=left}
    ]

        \node[note, font=\sffamily\normalsize] (inputs) {Inputs: OpenROAD repo, EDA literature $\mathcal{C}_{\mathrm{lit}}$.};

        \node[stage, below=0.2cm of inputs] (s0) {\textbf{S0: Repo graph \& docs}\\
            Build cross-language property graph $G$ and doc cards $\mathcal{C}_{\mathrm{repo}}$.};

        \node[stage, below=0.4cm of s0] (s1) {\textbf{S1: Literature-grounded planning}\\
            DSPy + RAG over $\mathcal{C}_{\mathrm{repo}}$ and $\mathcal{C}_{\mathrm{lit}}$ to emit high-level plans.};

        \node[stage, below=0.4cm of s1] (s2) {\textbf{S2: Localization}\\
            Map plans to concrete edit surfaces and granular steps \\
            $(\Delta_i, \text{tests, probes})$.};

        \node[stage, below=0.4cm of s2] (s3) {\textbf{S3: Autonomous execution}\\
            Agent applies diffs, runs flows, and gates on QoR \\
            QoR = (WL, ECP, DRC, timing).};

        \draw[arrow] (s0) -- (s1);
        \draw[arrow] (s1) -- (s2);
        \draw[arrow] (s2) -- (s3);

        \node[note, below=0.2cm of s3, font=\sffamily\normalsize] (outputs) {Outputs: Accepted code diffs and QoR improvements.};

        \draw[arrow, dashed]
            (s3.west) to[bend left=45]
            node[left, note]{\normalsize QoR}
            (s1.west);

    \end{tikzpicture}
    \vspace{-0.3in}
    \caption{The AuDoPEDA pipeline.}
    \label{fig:architecture_flow}
\vspace{-0.2in}
\end{figure}
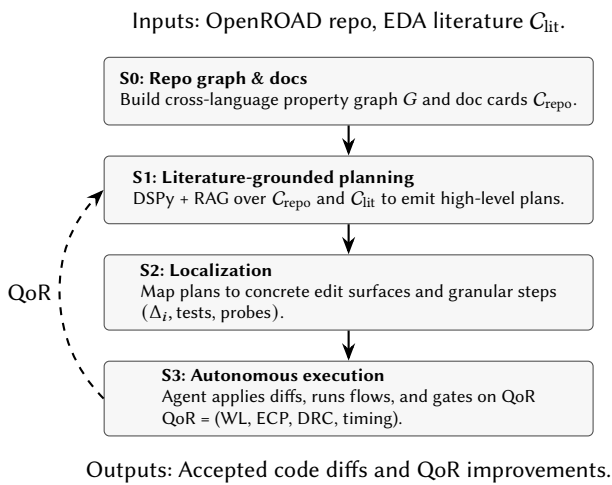

In this work, we ask whether an LLM-driven system can make \emph{autonomous,
repository-scale coding contributions} to a production EDA stack while closing the 
loop on quality-of-results (QoR). Our answer is a programmatic pipeline that 
(i) bootstraps machine-usable documentation and a code graph for OpenROAD,
(ii) composes domain literature with repository structure to synthesize plans,
and (iii) executes those plans via an agentic code editor that proposes diffs, compiles, 
runs flows, and hill-climbs against PPA targets (e.g., routed wirelength, effective 
clock period). Our thesis is simple: unlike generic software engineering, EDA is a highly 
specialized technical domain in which expertise is acquired by reading code, papers, 
manuals, and tool/flow documentation. There is no fundamental reason why an autonomous coding 
agent cannot be ``onboarded'' in the same way, via a documentation-first approach. 
Concretely, our system factors this task into four coupled stages:

\smallskip \noindent
\textbf{Stage 0: Documentation bootstrap and code graph.}
We parse the OpenROAD repository with \emph{tree-sitter} to obtain language-agnostic abstract syntax trees (ASTs) and structural edges (calls, includes, bindings), and compile them into a 
sparsified dependency directed acyclic graph (DAG) at file and function granularity \cite{treesitter}. A 
documentation generator (``Docmaker'') traverses this DAG bottom-up to produce concise, leaf-to-module cards capturing APIs, invariants, and conditions both before and after, thereby creating a hierarchical knowledge base suitable for downstream retrieval and planning. This staged, graph-aware summarization mitigates the context dilution typical of single-shot prompts over raw files \cite{hcgs,lewis2020rag}.

\smallskip \noindent
\textbf{Stage 1: Literature-grounded, DSPy-programmed planning.}
We treat planning as a declarative LLM program compiled by \emph{DSPy} \cite{dspy}. The planner retrieves from (a) the Docmaker cards and code graph, and (b) a domain corpus covering placement, clock tree synthesis (CTS), routing, and timing closure via retrieval-augmented generation \cite{lewis2020rag}. It emits \emph{high-level research directions} (e.g., ``reduce half-perimeter wirelength by adjusting legalization cost and congestion penalties''), each accompanied by acceptance tests, telemetry hooks, and rollback conditions that make the plans directly executable and evaluable.

\smallskip \noindent
\textbf{Stage 2: Plan localization to source.}
High-level plans are then projected onto concrete edit surfaces by aligning objectives with neighborhoods of the code graph (modules/files) and by extracting the relevant APIs and invariants from the leaf-level documentation cards. The result is a \emph{granular plan}: an ordered set of diffs annotated with pre-flight checks (build, unit, flow-smoke tests), runtime probes (timing, design rule check (DRC), and routing metrics), and post-conditions tied to QoR deltas. This bridges literature-informed intent and concrete, verifiable changes to the OpenROAD codebase and scripts.

\smallskip \noindent
\textbf{Stage 3: Autonomous execution with verification loops.}
A Codex-class coding agent, operating in a planner–executor architecture, applies the proposed diffs, compiles the modified code, and runs instrumented OpenROAD flows on designated benchmarks. The agent iterates using self-measured signals (e.g., routed wirelength (rWL), worst negative slack (WNS), total negative slack (TNS), via count and power) together with guardrails such as revert-on-regression and bisect-on-failure. It leverages tool-use patterns (code search, syntax trees, build logs) similar to SWE-agent \cite{sweagent}, and can opportunistically invoke tool APIs learned from examples (cf. Toolformer) \cite{toolformer}. Failures (compile errors, runtime crashes, QoR regressions) are converted into counterexamples and fed back into the granular plan for repair, closing a self-correcting feedback loop.

\smallskip \noindent
\textbf{Contributions.}
We introduce \textbf{AuDoPEDA}, the first \textbf{Au}tonomous \textbf{Do}cumentation and \textbf{P}lanning system for \textbf{EDA} codebases that integrates learned program synthesis with design-automation workflows. Our concrete contributions are:
\begin{itemize}
  \item \textbf{A graph-structured documentation maker} for OpenROAD, built with \textit{tree-sitter}, that constructs leaf-to-module cards capturing the invariants and interfaces required for downstream planning \cite{treesitter,hcgs}. This documentation pipeline \textbf{fully generalizes} to any EDA repository.
  \item \textbf{A literature-grounded planning layer}, implemented with \textit{DSPy}, which composes contextual documentation with EDA papers and wikis to generate testable, repository-aware research directions \cite{dspy,lewis2020rag}.
  \item \textbf{An autonomous executor for EDA codebases} that translates plans into code diffs, compiles, runs full design flows, and hill-climbs on metrics under safety constraints \cite{sweagent}. We intentionally scaffold atop the leading \textbf{Codex} agent from OpenAI, which is open-source and allows such modifications.
  \item \textbf{End-to-end validation on OpenROAD}, demonstrating that the system can produce accepted patches and measurable QoR improvements --- specifically, in routed wirelength, effective clock period, and total power consumption --- across fixed benchmarks and acceptance tests.
\end{itemize}

\smallskip \noindent
\textbf{Scope and positioning.}
Unlike prior work that either documents code \emph{or} assists local edits, our 
system couples (i) graph-structured documentation, (ii) literature-grounded 
planning, and (iii) agentic execution with QoR feedback, all tailored to 
an industrial-scale EDA repository. The design emphasizes \emph{programming} 
LLMs rather than ad hoc prompt engineering: each stage is a typed operator 
with explicit inputs and outputs (graphs, cards, plans, diffs, metrics), 
enabling determinism, auditability, and reuse across flows \cite{dspy}. 
Below, Section~\ref{sec:relatedwork} reviews relevant literature,
and Section~\ref{sec:methods} presents the pipeline end-to-end. 
Section~\ref{sec:exps} gives full-system results: documentation artifacts, 
high-level and granular implementation plans, and resulting code diffs, 
along with QoR (routed wirelength, effective clock period, and total power) improvements
obtained by applying these diffs to OpenROAD. Section~\ref{sec:conc} concludes 
with future directions for autonomous LLM agents in EDA.

\section{Related Work}
\label{sec:relatedwork}
Our work intersects four research vectors: the open-source EDA ecosystem, autonomous agents for software engineering, advanced methods for code representation and contextualization, and the application of LLMs to hardware design automation.

\smallskip \noindent
\textbf{The Open-Source EDA Ecosystem.}
The OpenROAD project provides a foundational platform for RTL-to-GDS implementation, targeting 24-hour, no-human-in-the-loop flows \cite{ajayi2019openroad,openroad}. This ecosystem comprises specialized engines interconnected via standardized interfaces. In placement, RePlAce implements routability-aware analytical global placement using non-linear optimization and density smoothing \cite{replace-tcad}, while DREAMPlace leverages GPU parallelism to accelerate wirelength and density objectives via deep learning frameworks \cite{dreamplace-dac}. For routing, FastRoute provides efficient congestion- and via-aware global routing heuristics \cite{fastroute-iccad,fastroute-2012}, and TritonRoute offers a detailed routing framework capable of achieving near-DRC-clean solutions on advanced benchmarks \cite{tritonroute-tcad}. OpenSTA supplies a Tcl-driven static timing analysis engine \cite{opensta-github}. 
At the flow orchestration level, OpenLane and OpenROAD-flow-scripts (ORFS) compose these tools into reproducible RTL-to-GDS pipelines \cite{openlane-woset,openlane-iccad, openroad-flow-scripts}. These works establish the necessary infrastructure—standardized metrics, verifiable outputs, and scriptable interfaces—for closed-loop evaluation. The heterogeneity (C++, Tcl, Python) and complexity of these large codebases present significant barriers for both human and autonomous modification, which our work addresses through structured documentation (S0) and graph-aware planning (S1/S2).

\smallskip \noindent
\textbf{Autonomous Agents and Repository-Scale Coding.}
The evolution of LLMs trained on code, from Codex \cite{chen2021codex} to models like CodeT5+ \cite{codet5plus} and CodeLlama \cite{roziere2023codellama}, has demonstrated strong performance on localized tasks. However, real-world engineering requires reasoning across large repositories, managing dependencies, and validating changes. SWE-bench formalized this challenge by benchmarking agents on resolving real-world GitHub issues with executable tests \cite{swebench}. To address this complexity, specialized agent architectures have emerged. SWE-agent demonstrated that optimized Agent-Computer Interfaces (ACIs) significantly improve performance by enhancing repository navigation, editing, and test execution \cite{sweagent}. AutoCodeRover further integrates sophisticated code search and navigation heuristics for autonomous program improvement \cite{zhang2024autocoderover}. Agentic systems often employ advanced strategies for planning, self-correction, and tool use. ReAct interleaves reasoning traces with actions to dynamically adjust plans \cite{yao2022react}. Reflexion introduced verbal reinforcement learning to iteratively refine agent trajectories based on environmental feedback \cite{shinn2023reflexion}. Toolformer showed that LMs can learn to invoke external APIs in a self-supervised manner \cite{toolformer}. To improve the reliability of these multi-step workflows, frameworks like DSPy treat LLM pipelines as declarative programs, compiling high-level objectives into optimized prompts and utilizing LM Assertions to enforce behavioral constraints \cite{dspy,dspy-assertions}. AuDoPEDA adopts these principles by compiling literature-grounded plans (via DSPy) and executing them under strict QoR feedback loops, extending the validation criteria from unit tests to physical design metrics.

\smallskip \noindent
\textbf{Code Representation and Contextualization.}
Effective repository scale coding relies on representing and retrieving relevant context. Early work on code representation learning, such as CodeBERT \cite{feng2020codebert}, applied BERT-style pre-training to bimodal (code and natural language) data. GraphCodeBERT advanced this by incorporating structural information, specifically data flow, into the pre-trained representation \cite{guo2021graphcodebert}. To handle extensive repository context, Retrieval-Augmented Generation (RAG) is standard~\cite{lewis2020rag}. Naive lexical retrieval often fails to capture code-level structural dependence. Recent work focuses on improving RAG by leveraging code structure. RepoCoder employs iterative retrieval and generation for repository-level completion \cite{zhang2023repocoder}. Advanced techniques utilize language-agnostic parsing (e.g., \emph{tree-sitter} \cite{treesitter}) to build dependency graphs, enabling hierarchical graph-based summarization that enhances context retrieval by respecting the code's inherent structure \cite{hcgs}. We employ this to build comprehensive, cross-language property graph and hierarchical documentation, allowing precise, structure-aware retrieval during planning (S1) and localization (S2).

\smallskip \noindent
\textbf{LLMs in Electronic Design Automation.}
Application of LLMs to EDA is rapidly expanding across the design flow \cite{llm-eda-survey}. 
In front-end design, ChipNeMo \cite{liu2023chipnemo}, RTLCoder \cite{wang2024rtlcoder}, etc. 
use LLMs for automated RTL generation, optimization, and conversational engineering assistance.
In the back-end, LLMs are increasingly used for optimization and verification. While traditional 
ML/RL methods have been used for parameter tuning (e.g., AutoTuner \cite{jung2021metrics2}), 
LLMs offer new interaction paradigms. EDA Corpus provides a dataset for training LLMs to 
interact with OpenROAD scripts and analyze flows \cite{eda-corpus}. In physical verification, 
DRC-Coder employs an agentic system to decompose DRC rule interpretation and synthesize 
executable verification code \cite{drc-coder-arxiv}.

Unlike prior work focused on RTL generation, scripting assistance or parameter tuning, 
AuDoPEDA targets \emph{autonomous code modification within the core C++/Tcl PD stack itself}. 
It uniquely couples graph-structured documentation, literature-grounded planning, and agentic execution with closed-loop PPA evaluation, enabling an agent to directly improve underlying algorithms and configurations of an industrial-scale EDA repository.

\vspace*{-1em}

\section{Methodology}
\label{sec:methods}

Our system, AuDoPEDA, decomposes the task of autonomous EDA code contribution into four structured stages: (S0) repository graphing and documentation bootstrap, (S1) literature-grounded planning, (S2) plan localization to code, and (S3) autonomous execution with QoR feedback. Each stage consumes and produces versioned artifacts (graphs, cards, plans, diffs, metrics), ensuring auditability and deterministic replay. We target the OpenROAD stack \cite{openroad,openroad-github} and its components (e.g., RePlAce, FastRoute, TritonRoute, OpenSTA) \cite{replace-tcad,fastroute-iccad,tritonroute-tcad,opensta-github}. An overall view appears in Figure~\ref{fig:architecture_flow}.

\subsection{S0: Repository Graphing and Documentation Bootstrap}

\smallskip \noindent
\textbf{Goal.} To make a large, multi-language EDA codebase (like OpenROAD) understandable to an autonomous agent, we first construct a detailed map of the code—a property graph $G$—and generate machine-usable documentation cards. These artifacts capture the code's structure, interfaces, and invariants, enabling downstream planning.

\smallskip \noindent
\textbf{S0.1 Parsing and Graph Construction.}
We analyze the OpenROAD repository, covering C/C++ (core engines), Tcl (scripts), Python (utilities), and Verilog (design inputs).
\begin{itemize}
  \item \textbf{Language-Agnostic Parsing.} We use \emph{tree-sitter} \cite{treesitter} to parse these languages into ASTs. We incorporate build information (from CMake) to accurately resolve dependencies and include paths, ensuring the parser understands the code as the compiler does.
  \item \textbf{The Code Graph ($G$).} We translate the ASTs into a unified graph representation. Nodes in $G$ represent key code elements: Files, Declarations (types, functions, classes), Definitions (implementations), and Callsites.
  \item \textbf{Typed Edges.} We connect these nodes with typed edges that capture relationships, such as \texttt{calls}, \texttt{includes}/\texttt{imports}, and variable \texttt{bindings}.
\end{itemize}

\smallskip \noindent
\textbf{S0.2 Cross-Language Linking and Simplification.}
A crucial step is linking the scripting interface (Tcl/Python) to the C++ core. We identify where Tcl commands are registered in C++ (e.g., using \texttt{Tcl\_CreateCommand} wrappers) and add \texttt{script\_invokes} edges. This allows the agent to trace execution from a high-level script command down to the engine implementation.

To manage complexity, we simplify the graph:
\begin{itemize}
    \item \textbf{Condensation.} We condense tightly coupled groups of functions (Strongly Connected Components) into single nodes, reducing noise from internal utility calls.
    \item \textbf{Sparsification and Filtering.} We prune less informative edges in dense areas and exclude non-essential code, such as third-party libraries and test-only files.
\end{itemize}

\smallskip \noindent
\textbf{S0.3 Docmaker Pipeline and Documentation Cards.}
The ``Docmaker'' automatically generates documentation cards by traversing the graph $G$ bottom-up (from utilities to main entry points).

\begin{itemize}
  \item \textbf{Evidence Extraction.} For each node, we extract key information: function signatures, default parameters, assertions, configuration flags (Tcl options), error messages, and neighborhood context.
  \item \textbf{Abstractive Summarization.} A code-specialized LLM converts this evidence into a concise summary card with standardized sections: \texttt{Role}, \texttt{Inputs/Outputs}, \texttt{Pre/Postconditions}, and \texttt{Config Knobs}.
  \item \textbf{Validation.} We validate the generated cards for accuracy: checking that all referenced APIs exist in $G$ and that the summary matches the code's type signatures. This ensures the documentation is reliable for the agent.
\end{itemize}

\smallskip \noindent
\textbf{S0.4 Indexing and Retrieval.}
To enable efficient search, we index cards and code snippets using traditional sparse retrieval (BM25) and modern dense embeddings. This is combined with structural information from the graph, prioritizing results that are contextually relevant \cite{hcgs}. The pipeline is incremental, updating only affected parts of the graph when the codebase changes.

\subsection{S1: Literature-Grounded Planning (DSPy)}
\textbf{Goal.} Generate high-level, executable research plans combining knowledge of OpenROAD (from S0) with EDA literature.

\smallskip \noindent
\textbf{Knowledge Corpora.} We maintain two corpora for retrieval: $\mathcal{C}_{\mathrm{repo}}$ (the Docmaker cards from S0) and $\mathcal{C}_{\mathrm{lit}}$ (EDA papers, tutorials, and documentation covering placement, routing, timing, etc.). Documents are domain-tagged (e.g. \texttt{gp}, \texttt{sta}) for retrieval.

\smallskip \noindent
\textbf{DSPy-Programmed Planner.} We treat planning not as a single prompt, but as a declarative LLM program compiled using \emph{DSPy} \cite{dspy}. This approach provides structure and reliability. The program follows a Retrieve-Synthesize-Validate structure:

\begin{enumerate}
  \item \textbf{Retrieve.} Given an objective (e.g., \emph{reduce routed wirelength}), the planner retrieves relevant information from both $\mathcal{C}_{\mathrm{repo}}$ and $\mathcal{C}_{\mathrm{lit}}$ \cite{lewis2020rag}. A re-ranker ensures the selected documents offer diverse perspectives and align with the repository structure.
  \item \textbf{Synthesize.} The planner integrates this information for a \emph{high-level plan} - namely the \texttt{objective}, \texttt{hypotheses} (derived from literature), proposed \texttt{interventions} (e.g., adjusting cost models, tuning knobs), and \texttt{telemetry} (tracked metrics). It also suggests potential code locations and APIs for the interventions.
  \item \textbf{Validate.} The plan is checked against predefined constraints (LM Assertions) and the code graph $G$. This ensures that proposed interventions are feasible (e.g., parameters are within range, referenced APIs exist, core invariants are respected) before execution.
\end{enumerate}

\subsection{S2: Localization and Granular Plans}
\textbf{Goal.} Translate the high-level plan from S1 into a concrete, ordered sequence of code edits and verification steps.

\smallskip \noindent
\textbf{Edit-Surface Localization.} We identify the specific locations (files, functions, scripts) in the codebase where changes are needed by mapping the interventions from S1 onto the code graph $G$. We aim to find the minimal set of nodes that cover the required APIs and passes, while also minimizing the ``blast radius'' — the potential impact on dependent modules, assessed using graph centrality and historical change frequency.

\smallskip \noindent
\textbf{Granular Plan Representation.} The output is a \emph{granular plan} detailing the implementation sequence. Each step includes:
\begin{itemize}
    \item $\Delta_i$: The intended code modification (diff intent).
    \item \texttt{Pre}$_i$: Pre-run checks (e.g., build health, unit tests, formatting).
    \item \texttt{Run}$_i$: The specific OpenROAD flow configuration to execute (design, process design kit (PDK), parameters).
    \item \texttt{Probe}$_i$: The metrics to monitor (QoR counters, logs).
    \item \texttt{Post}$_i$: Acceptance criteria (e.g., QoR improvement targets) and rollback conditions if the change causes regressions.
\end{itemize}
This structure ensures that the plan is verifiable, deterministic, and safe to execute autonomously.

\subsection{S3: Autonomous Execution with QoR Feedback}
\textbf{Goal.} Execute the granular plan, apply code modifications, run EDA flows, and improve the results based on measured PPA.

\smallskip \noindent
\textbf{Agent Architecture and Tools.} We employ a Codex-class coding agent structured with a planner–executor interface, similar to SWE-agent \cite{sweagent}. The executor utilizes a specialized tool palette for interacting with the EDA environment:
\begin{itemize}
  \item \texttt{code\_edit}: Apply structured edits directly at the AST level (using tree-sitter coordinates), ensuring precise modifications.
  \item \texttt{build}: Compile OpenROAD and check unit tests.
  \item \texttt{flow}: Run non-interactive OpenROAD flows up to specific checkpoints (e.g., placement, routing, signoff static timing analysis (STA)).
  \item \texttt{measure}: Extract PPA from tool reports (e.g., routed wirelength/DRCs from TritonRoute; WNS, TNS, and total power from OpenSTA).
  \item \texttt{rollback} / \texttt{bisect}: Revert changes or isolate regressions.
\end{itemize}

\smallskip \noindent
\textbf{QoR Gating and Metric Model.} We evaluate candidate changes using a composite score derived from normalized metrics (wirelength (WL), via count, density, timing). Crucially, we employ hard \emph{gates}: a change is immediately rejected if it introduces new DRCs, worsens WNS beyond a predefined threshold, or causes build/test failures. A candidate is accepted only if the composite score improves and all gates pass.

\smallskip \noindent
\textbf{Search and Iteration Policy.} The agent operates in a closed feedback loop. It alternates between (a) \emph{local repair} (fixing build errors or test failures introduced by a diff) and (b) \emph{QoR hill-climbing}. During hill-climbing, the agent proposes several diffs, evaluates them using fast proxy flows (e.g., place + global-route), and promotes the most promising candidates to full signoff runs. The best non-regressing change is committed. Failures are fed back to the planner (S1/S2) to refine future attempts.

\subsection{Evaluation Protocol and Implementation}

\smallskip \noindent
\textbf{Experimental Setup.} We evaluate the system using standard open designs on public PDKs (e.g., ASAP7, SKY130HD, Nangate45) with a fixed OpenROAD snapshot. We use both short proxy flows for quick evaluation and full signoff flows for final validation. Builds are containerized for reproducibility, and all artifacts (plans, diffs, logs, QoR reports) are versioned. Our protocol has been primarily validated on a GCP-Codex stack, though we anticipate generalization. Our Apptainer environment, pre-built binaries, and source files are available at \cite{this_work-DockerHub}.

\smallskip \noindent
\textbf{Safety and Rollback.} The system employs two layers of safety guards. (i) \emph{Static guards} in S1/S2 check API existence and respect invariants before execution. (ii) \emph{Runtime guards} in S3 enforce build, test, DRC, and timing gates. Violation triggers an automatic rollback, and the failure is recorded as a counterexample to inform future planning attempts. We cache intermediate flow results (e.g., placed DEF) to accelerate the hill-climbing loop.

\section{Experiments}
\label{sec:exps}
We present some elements in order. \textbf{First}: The existence of the docs which power our autonomous diff-applying agent. \textbf{Second}: The existence of high-level plans, which interface these docs with research papers. To illustrate our point, research papers are chosen solely from the proceedings of prominent EDA conferences (i.e., DAC, ICCAD). In terms of modification targets, we only allow modifications to \textbf{DPL, GPL, RSZ} --- i.e., \textbf{Detailed Placement, Global Placement, Resizer} of OpenROAD's setup. \textbf{Third}: The resultant product creates granular implementation plans, which are usable as fuel for the coding agent pipeline. \textbf{Fourth}: The coding agent applies the diffs itself, resulting in improved QoR. All experiments are run on a Google Cloud Machine with 112 vCPUs under an AMD EPYC Hypervisor setup, 220 GB RAM. Note that in general, our results apply to x86 architectures and that
ARM64 compatibility is unclear -- a situation that holds for OpenROAD and ORFS in general. 
The baseline ORFS hash for benchmarking QoR is 93c42b, which associates to the
OpenROAD hash 7bc521.\footnote{The commit hashes of submodules such as yosys and yosys-slang are synchronized automatically. For detailed  information and source code, see
\cite{this_work-DockerHub}.}

\vspace*{-0.21em}

\subsection{Improvement of Wirelength}
To reduce final routed wirelength, we let the agent restructure the detailed placement 
global swapper into a two-pass, routability-driven engine. A baseline half-perimeter wirelength (HPWL) estimate is established in a brief profiling pass, and is then used
to govern a budgeted optimization. The given budget allows the placer to accept 
small, controlled deviations from the initial wirelength estimate if doing so yields 
a significant improvement in routability. This prevents the optimizer from 
creating placements that are locally optimal in terms of wirelength estimation but 
globally difficult to route, which would ultimately inflate rWL. We show
results on aes, ibex and jpeg circuits under the ASAP7, SKY130HD and Nangate45 PDKs,
along with four macro-heavy Nangate45 circuits (ariane133, ariane136, 
bp\_fe, swerv\_wrapper). These benchmark designs are commonly used in the field~\cite{jung2021metrics2}.

At the core of the engine is a cost function that balances the
placement wirelength estimate against a site-level utilization
density metric. This metric combines cell area and pin count
within each placement region: the area component identifies
physically congested areas where routing resources are scarce,
while the pin component captures regions with high routing
demand where many nets converge. The swapper penalizes moves
that would exacerbate congestion and disperses cells from
high-density regions, thus preserving track availability and enabling
more direct net routing paths.

A hybrid move generation policy combines deterministic, wirelength-optimal proposals with stochastic, exploratory moves to effectively navigate the solution space. The decision to accept a move is guided by a data-driven \emph{adaptive weight}, which enables a principled tradeoff between the wirelength cost and routability benefit. This weight is learned at runtime by sampling a set of trial swaps to find the average magnitude of change for both the wirelength estimate and the routability score. The weight then normalizes the abstract routability metric into the same physical units as wirelength. This ensures that the algorithm only spends its wirelength budget on moves that provide a meaningful improvement in placement routability. By proactively mitigating congestion and preventing routing hotspots, this methodology produces a placement that is inherently easier to wire, leading to consistent and measurable reductions in final rWL. The improvements  do not arise from per-design parameter tuning, but from a single, repository-level modification discovered by the agent and exposed as a flow variant. 
Under high core utilization, which is the regime where WL increases due to detouring
and routing failures are more common, the added mode (i) remains localized and safe 
to toggle, (ii) measurably improves routed WL in a multi-PDK setting, and 
(iii) achieves a nuanced tradeoff between WL and routability when congestion makes 
pure WL minimization undesirable. 
Table~\ref{tab:highutil} shows the flow configurations that we use to 
illuminate this.\footnote{The terms used in this table correspond to parameters 
in ORFS. Specifically, Util refers to \texttt{CORE\_UTIL}, LB add-on to \texttt{PLACEMENT\_LB\_ADDON}, Aspect to \texttt{CORE\_ASPECT\_RATIO}, and Margin to \texttt{CORE\_MARGIN}. In addition, DPO enable denotes \texttt{ENABLE\_DPO=1}, and EQY disable denotes \texttt{EQUIVALENCE\_CHECK=0}. Finally, we apply \texttt{undefine CORE\_AREA} and \texttt{undefine DIE\_AREA} to all designs so that the floorplan is determined based on \texttt{CORE\_UTIL}. See \cite{this_work-DockerHub}.} 
Table~\ref{tab:wl} summarizes the results. Utilization values in Table~\ref{tab:highutil} are arrived at by incrementing core utilization upwards in steps of $5$(\%) until routability failure occurs.

\begin{table}[h]
\vspace{-0.05in}
\centering
\footnotesize
\setlength{\tabcolsep}{4pt} 
\resizebox{\linewidth}{!}{%
\begin{tabular}{l l c c c}
\toprule
\textbf{Process} & \textbf{Design} & \textbf{Util} & \textbf{LB add-on} & \textbf{Aspect (Margin)} \\
\midrule
ASAP7 & aes  & 75\% & 0.2 & -- \\
ASAP7 & ibex & 70\% & 0.2 & -- \\
ASAP7 & jpeg & 70\% & 0.2 & -- \\
\midrule
SKY130HD & aes  & 30\% & 0.2 & -- \\
SKY130HD & ibex & 50\% & 0.2 & -- \\
SKY130HD & jpeg & 60\% & 0.2 & -- \\
\midrule
Nangate45 & aes  & 85\% & 0.2 & -- \\
Nangate45 & ibex & 30\% & 0.2 & -- \\
Nangate45 & jpeg & 30\% & 0.2 & -- \\
Nangate45 & bp\_fe & 30\% & 0.11 & -- \\
Nangate45 & ariane133 & 30\% & -- & 1(5) \\
Nangate45 & ariane136 & 30\% & -- & -- \\
Nangate45 & swerv\_wrapper & 30\% & 0.08 & 1(5) \\
\bottomrule
\end{tabular}%
}
\caption{Flow configurations for wirelength improvement testing, all with DPO 
enable and EQY disable.}
\label{tab:highutil}

\centering
\resizebox{\linewidth}{!}{%
\begin{tabular}{l l r r r}
\toprule
\textbf{Platform} & \textbf{Design} & \textbf{Base rWL} & \textbf{Our rWL} & \textbf{$\Delta$rWL (\%)} \\
\midrule
ASAP7    & aes      & 64{,}640   & 62{,}710    & $-2.99$ \\
ASAP7    & ibex     & 80{,}402   & 80{,}823    & $+0.52$ \\
ASAP7    & jpeg     & 154{,}484  & 152{,}232   & $-1.46$ \\
\midrule
SKY130HD & aes      & 659{,}778  & 633{,}899   & $-3.92$ \\
SKY130HD & ibex      & 646{,}855  & 643{,}006   & $-0.60$ \\
SKY130HD & jpeg     & N/A        & 1{,}201{,}778 & N/A \\
\midrule
Nangate45  &   aes          & 230{,}044   & 217{,}415   & $-5.49$ \\
Nangate45  &   ibex        & 248{,}641   & 248{,}429   & $-0.09$ \\
Nangate45  &   jpeg        & 565{,}979   & 554{,}902   & $-1.96$ \\
Nangate45  &   bp\_fe      & 1{,}603{,}884 & 1{,}634{,}916 & $+1.94$ \\
Nangate45  &   ariane133   & 7{,}831{,}361 & 7{,}523{,}708 & $-3.93$ \\
Nangate45  &   ariane136    & 7{,}986{,}048 & 7{,}509{,}944 & $-5.96$ \\
Nangate45  &   swerv\_wrapper & 4{,}310{,}916 & 4{,}239{,}837 & $-1.65$ \\
\bottomrule
\end{tabular}%
}
\caption{Routed wirelength (rWL, in units of $\mu$m) outcomes from wirelength
improvement testing.}
\label{tab:wl}
\end{table}


\noindent
The same autonomous diff is applied unchanged across all platforms and designs. Further, the gain appears to be a ``free lunch'' --- no discernible worsening was observed for effective clock period (ECP), power, instance area or instance count, with ECP and power showing an average \textbf{improvement} instead.

\subsection{Better ECP at Aggressive Timing}

For this experiment, we provide the system with the docs of \textbf{GPL} (Global 
Placement) and \textbf{RSZ} (Resizer), and ask it to generate a framework 
which lowers the ECP when target clock period (TCP) is tightened to $0.85 \times$ the 
default in ORFS. 

The agent's generated strategy autonomously develops a sophisticated, multi-pronged approach to address this aggressive timing target. Instead of just increasing timing pressure globally, the strategy focuses on surgical, stable interventions. Its architecture, derived from the provided documentation, is built on the following principal themes:

\begin{itemize}
    \item{\textbf{Weight-Intensity Control.} In timing-driven placement, each net is assigned a 
    slack-based weight. The agent provides a parameter that caps the maximum weight 
    magnitude, allowing the user to bound the force applied to the worst-slack nets. 
    When timing-driven mode is enabled, the flow collects net slacks, constructs a worst-net 
    set, and incorporates a predefined maximum scaling factor during weight computation. 
    Raising this upper bound increases timing-improvement pressure, while lowering it tends to keep placement more stable. This parameter therefore controls the aggressiveness of timing optimization.}
    \item{\textbf{Weight-Curve Shaping.} The agent normalizes slack values and applies a nonlinear transformation before incorporating them into the weight calculation. A steeper curve concentrates weight more heavily on the worst-slack nets, whereas a gentler curve spreads weight more broadly. This provides control over the shape of the weight distribution.}
    \item{\textbf{Reference-Point Redefinition.} The agent allows the normalization reference to be set to 0-slack, thereby avoiding excessive emphasis on nets that already exhibit positive slack. After constructing the worst-net set, each net’s slack is normalized relative to the 0-slack reference and used to compute the net's weight. As a result, nets with sufficient margin naturally exert less influence, while violating or near-critical nets receive proportionally higher weight.}
    \item{\textbf{Focus-Range Adjustment.} When forming the worst-net set, the targeted scope can be constrained by a percentage. Nets are sorted by slack; only the top subset is retained for weight computation, while the remaining nets are excluded. Narrowing the scope concentrates optimization on the most severe nets, whereas widening it expands the set of nets prioritized for improvement. This provides explicit control over the breadth of timing-driven interventions.}
    \item{\textbf{Length-Based Adjustment.} Slack alone may not fully reflect the structural disadvantage of long nets. The agent incorporates a net’s bounding-box length (HPWL-like length) computed during placement into the weight calculation. After computing the slack-based weight, the flow computes the ratio of the net length to the global average length and uses this ratio to adjust the weight. Consequently, longer nets tend to receive higher weights, increasing their priority for timing improvement.}
    \item{\textbf{Control of Length Contribution.} The contribution of the length-based adjustment is configurable. With a low setting, slack-based weighting dominates; with a high setting, length has a stronger influence. This enables fine-grained tuning of length bias to match design-specific characteristics.}
    \item{\textbf{Repair-Loop Convergence Control.} During timing repair, some passes may temporarily degrade slack. The agent allows the user to specify the maximum number of such degrading passes permitted. At each pass, slack improvement is evaluated. If the accumulated number of degrading passes exceeds a threshold, the flow restores the prior state and terminates. This helps manage convergence stability and runtime cost.}
    \item{\textbf{Selection of Analysis Fidelity and Runtime.} Within RSZ, the accuracy and runtime of timing analysis can vary substantially depending on the parasitics model. The agent enables selection between a higher-fidelity model and a faster approximate model, allowing the tradeoff between analysis fidelity and runtime to be governed as an explicit policy.}
\end{itemize}

The combined effect is a more robust and targeted timing-driven flow that successfully reclaims slack from the tightened clock without sacrificing routability or hold timing. We test our strategy on 
larger Nangate45 circuits (details given in 
Table~\ref{tab:ng45metadata}\footnote{Incorrect wiring 
of macros in the ariane133 testcase at the ORFS repository causes one macro to be dropped 
from the netlist. We have raised a pull request \cite{ariane133_pr} to OpenROAD project maintainers to correct 
this.})
for which the target 
clock period is reduced by 15\% (50\% in the case of ariane136) from the 
original value in ORFS. As shown in Table~\ref{tab:freq}, this strategy, developed 
autonomously by our agent, achieves significant ECP reductions 
relative to Base OpenROAD.

\begin{table}[h!]
\centering
\setlength{\tabcolsep}{4pt}

\begin{tabular}{l r r r r}
\hline
\textbf{Design} & \textbf{\#Cells} & \textbf{\#Macros} & \textbf{\#Nets} & \textbf{\#Pins} \\
\hline
ariane133      & 184314 & 132 & 195662 & 620474 \\
ariane136      & 194547 & 136 & 205959 & 643350 \\
bp\_fe         & 38924  & 11 & 41418  & 114292 \\
swerv\_wrapper & 107466 & 28 & 113694 & 358017 \\
\hline
\end{tabular}
\caption{Attributes of larger Nangate45 circuits.}
\label{tab:ng45metadata}

\begin{tabular}{l r r r r}
\hline
\textbf{Design} & \textbf{TCP} & \textbf{Base ECP} & \textbf{Our ECP} & \textbf{$\Delta$ECP} \\
\hline
ariane133      & 3.4  & 3.59  & 3.43  & -4.6\% \\
ariane136      & 3.0  & 3.78  & 3.41  & -10.0\% \\
bp\_fe         & 1.53 & 1.71  & 1.65  & -3.4\% \\
swerv\_wrapper & 1.7  & 2.19  & 2.16  & -1.4\% \\
\hline
\end{tabular}
\caption{Timing results in Nangate45 (TCP/ECP in ns).}
\vspace{-0.2in}
\label{tab:freq}

\end{table}

\smallskip
\noindent
\textbf{Other aspects of interest.} Unlike choosing to create a \textbf{bespoke} code diff applicable per circuit, i.e., creating a diff specific to each circuit, the \textbf{same diff} can successfully achieve the results in Table~\ref{tab:freq}.
Further, despite creating changes in \textbf{RSZ} and \textbf{GPL} that by themselves cannot accomplish much, the system can successfully ensure that these modifications interact with each other in a positive-sum fashion -- an important criterion in 
the field of VLSI CAD, where many interlocking modules together comprise
a complete RTL-to-GDSII pipeline.

\subsection{Improvement of Power}
For this experiment, we provide the system with the RSZ documentation and ask it to reduce power by modifying the resizer module in ORFS.
The agent’s generated strategy autonomously develops a more advanced recovery flow than the baseline. Instead of only downsizing cells along slack-eligible timing paths, the strategy shifts to a design-wide instance view and applies staged recovery under timing safety checks.
Its architecture, derived from the provided documentation and reflected in the modified RSZ code, is built on the following principal themes:
First, it replaces path-by-path target selection with instance ranking based on power contribution and timing margin, then prioritizes high-power non-critical cells.
Second, it extends the recovery flow beyond sizing by combining buffer removal, cell downsizing, and low-leakage VT swapping in a multi-pass sequence.

\begin{table}[h]
\vspace{-0.1in}
\centering
\resizebox{\linewidth}{!}{%
\begin{tabular}{l l r r r r}
\toprule
\textbf{Platform} & \textbf{Design} & \textbf{Base Pwr} & \textbf{Our Pwr} & \textbf{$\Delta$Pwr (\%)} & \textbf{$\Delta$ECP (\%)} \\
\midrule
ASAP7    & aes      & 153.74  & 150.75  & $-1.947$   & 9.836 \\
ASAP7    & ibex     & 58.10   & 47.24   & $-18.703$  & 2.606 \\
ASAP7    & jpeg     & 119.70  & 117.48  & $-1.853$   & 2.880 \\
\midrule
SKY130HD & aes      & 456.94  & 437.55  & $-4.245$   & 3.529 \\
SKY130HD & ibex     & 93.20   & 75.14   & $-19.383$  & 4.433 \\
SKY130HD & jpeg     & 485.90  & 428.01  & $-11.914$  & 1.786 \\
\midrule
Nangate45  & aes    & 385.29  & 373.67  & $-3.017$   & 1.339 \\
Nangate45  & ibex   & 95.98   & 89.37   & $-6.883$   & 3.644 \\
Nangate45  & jpeg   & 499.22  & 454.71  & $-8.915$   & 5.647 \\
\bottomrule
\end{tabular}%
}
\caption{Power outcomes (Pwr = total power, in mW) from power improvement testing.}
\label{tab:pwr}
\vspace{-0.3in}
\end{table}

Table \ref{tab:pwr} summarizes the experimental results, showing power reductions in all cases.\footnote{All power values are extracted at the finish stage of the flow using OpenROAD \texttt{report\_power\_metric}, and the reported total power is the sum of internal, switching, and leakage power values. In addition, the \texttt{RECOVER\_POWER} variable is set to 100 in the flow configuration of each design. OpenROAD/OpenSTA power and timing analyses have been previously reported to show a reasonable level of correlation with commercial tools~\cite{RDF_ICCAD24}.} This improvement stems from the difference in optimization scope. A path-based method concentrates its effort on specific timing paths, so it can miss high-power cells that reside on non-critical paths. In contrast, the instance-based method ranks all cells by their power contribution and available slack, then directly targets high-impact, non-critical instances first. As a result, it achieves larger power savings under the same timing constraints. In addition, buffer removal and VT swapping provide extra leverage beyond simple downsizing, reducing both dynamic power (via switching activity and capacitance) and leakage power, which further broadens the room for improvement. The last column reports the change in ECP, which captures the overhead induced by the new recovery flow. As expected, we observe ECP increases in all benchmarks (by roughly 1–10\%), but this overhead is generally modest relative to the achieved power savings.

\vspace{0.0in}
\section{Conclusions}
\label{sec:conc}

We demonstrate that autonomous, repository-scale code modification for industrial-grade 
EDA flows is both feasible and effective when supported by structured documentation, 
literature-grounded planning, and agentic execution with QoR feedback.
AuDoPEDA tightly couples graph-based documentation, DSPy-programmed research planning, 
localized granular plans, and a Codex-class executor to form an end-to-end system 
capable of generating legitimate, verifiable improvements to the OpenROAD physical design stack.
Experiments across multiple PDKs, designs, and operating regimes show that the agent 
can (i) synthesize coherent research hypotheses, (ii) translate them into safe,
diff-level implementations, and (iii) achieve measurable PPA improvements without per-design 
tuning or manual intervention. These include nontrivial ECP reductions under aggressive 
timing, and substantial routed wirelength improvements in high-utilization, 
congestion-heavy configurations, and consistent power reductions across multiple PDKs and designs.
Importantly, all improvements are produced by a single 
autonomous diff per experiment category, underscoring AuDoPEDA’s ability to discover 
repository-scale, flow-consistent enhancements rather than one-off parameter tweaks.

Overall, AuDoPEDA suggests a new paradigm for EDA R\&D via agents that are not mere
assistants or script generators, but full participants that read code, propose algorithmic 
changes and validate their effects, and produce reproducible contributions. Future work 
includes integrating formal verification signals into the QoR gate, exploring multi-agent 
collaboration across flow stages, and extending the documentation and planning framework 
to additional open-source EDA systems. We believe that AuDoPEDA constitutes a first step 
toward self-improving, continuously learning design-automation toolchains.

\section*{Acknowledgments}
This work is partially supported by the Samsung AI Center.



\end{document}